\documentclass{emulateapj}

\usepackage{longtable}

\shorttitle{AGN Host Morphologies at $z\sim2$}
\shortauthors{Kocevski et al.}

\begin{document}

\title{CANDELS: Constraining the AGN-Merger Connection with Host Morphologies at \lowercase{$z\sim2$}}

\author{Dale D.~Kocevski\altaffilmark{1}, 
S.~M.~Faber\altaffilmark{1}, 
Mark Mozena\altaffilmark{1}, 
Anton M.~Koekemoer\altaffilmark{2}, 
Kirpal Nandra\altaffilmark{3}, 
Cyprian Rangel\altaffilmark{4}, 
Elise S.~Laird\altaffilmark{4}, 
Marcella Brusa\altaffilmark{3}, 
Stijn Wuyts\altaffilmark{3}, 
Jonathan R.~Trump\altaffilmark{1}, 
David C.~Koo\altaffilmark{1}, 
Rachel S.~Somerville\altaffilmark{2,5}, 
Eric F.~Bell\altaffilmark{6}, 
Jennifer M.~Lotz\altaffilmark{2}, 
David M Alexander\altaffilmark{7}, 
Frederic Bournaud\altaffilmark{8}, 
Christopher J.~Conselice\altaffilmark{9}, 
Tomas Dahlen\altaffilmark{2}, 
Avashi Dekel\altaffilmark{10}, 
Jennifer L.~Donley\altaffilmark{2},
James S.~Dunlop\altaffilmark{11}, 
Alexis Finoguenov\altaffilmark{3}\altaffilmark{12}, 
Antonis Georgakakis\altaffilmark{13}
Mauro Giavalisco\altaffilmark{14}, 
Yicheng Guo\altaffilmark{14}, 
Norman A.~Grogin\altaffilmark{2}, 
Nimish P.~Hathi\altaffilmark{15}, 
St\'ephanie Juneau\altaffilmark{16}, 
Jeyhan S.~Kartaltepe\altaffilmark{17}, 
Ray A.~Lucas\altaffilmark{2}, 
Elizabeth J.~McGrath\altaffilmark{1},
Daniel H.~McIntosh\altaffilmark{18}, 
Bahram Mobasher\altaffilmark{19}, 
Aday R.~Robaina\altaffilmark{20}, 
David Rosario\altaffilmark{3}, 
Amber N.~Straughn\altaffilmark{21}, 
Arjen van der Wel\altaffilmark{3}, 
Carolin Villforth\altaffilmark{2}}

\affil{University of California Observatories/Lick Observatory, University of
  California, Santa Cruz, CA 95064}
\altaffiltext{1}{University of California, Santa Cruz}
\altaffiltext{2}{Space Telescope Science Institute}
\altaffiltext{3}{Max-Planck-Institut f\"ur Extraterrestrische Physik}
\altaffiltext{4}{Imperial College of Science}
\altaffiltext{5}{Johns Hopkins University}
\altaffiltext{6}{University of Michigan}
\altaffiltext{7}{Durham University}
\altaffiltext{8}{Laboratoire AIM Paris-Saclay}
\altaffiltext{9}{University of Nottingham}
\altaffiltext{10}{The Hebrew University}
\altaffiltext{11}{Royal Observatory, University of Edinburgh}
\altaffiltext{12}{University of Maryland}
\altaffiltext{13}{National Observatory of Athens}
\altaffiltext{14}{University of Massachusetts}
\altaffiltext{15}{Observatories of the Carnegie Institution of Washington}
\altaffiltext{16}{Steward Observatory, University of Arizona}
\altaffiltext{17}{National Optical Astronomy Observatories}
\altaffiltext{18}{University of Missouri}
\altaffiltext{19}{University of California, Riverside}
\altaffiltext{20}{Institut de Ciencies del Cosmos}
\altaffiltext{21}{NASA Goddard Space Flight Center}

\email{kocevski@ucolick.org}

\begin{abstract}

Using \emph{HST}/WFC3 imaging taken as part of the Cosmic Assembly Near-infrared
Deep Extragalactic Legacy Survey (CANDELS), we examine the role that major
galaxy mergers play in triggering active galactic nuclei (AGN) activity at
$z\sim2$.  Our sample consists of 72 moderate-luminosity ($L_{\rm X} \sim 10^{42 - 44}$
erg s$^{-1}$) AGN at $1.5<z<2.5$ that are selected using the 4 Msec
\emph{Chandra} observations in the \emph{Chandra} Deep Field South, the
deepest X-ray observations to date.  Employing 
visual classifications, we have analyzed the rest-frame optical
morphologies of the AGN host galaxies and compared them to a mass-matched
control sample of 216 non-active galaxies at the same redshift.
We find that most of the AGN reside in disk galaxies
($51.4^{+5.8}_{-5.9}\%$), while a smaller percentage are found in spheroids ($27.8^{+5.8}_{-4.6}\%$).
Roughly $16.7^{+5.3}_{-3.5}\%$ of the AGN hosts have highly
disturbed morphologies and appear to be involved in a major merger or
interaction, while most of the hosts ($55.6^{+5.6}_{-5.9}\%$) appear relatively
relaxed and undisturbed.  These fractions are statistically consistent with
the fraction of control galaxies that show similar morphological
disturbances.  These results suggest that the hosts of moderate-luminosity AGN
are no more likely to be involved in an ongoing merger or interaction
relative to non-active galaxies of similar mass at $z\sim2$.  
The high disk fraction observed among the AGN hosts also appears to be at odds
with predictions that merger-driven accretion should be the dominant AGN
fueling mode at $z\sim2$, even at moderate X-ray luminosities.  
Although we cannot rule out that minor mergers are responsible for triggering
these systems, the presence of a large population of relatively undisturbed
disk-like hosts suggests that secular processes 
play a greater role in fueling AGN activity at $z\sim2$ than previously thought.

\end{abstract}


\section{Introduction}

Although it has been established that super-massive black holes (SMBHs) lie
at the center of most, if not all, massive galaxies (Magorrian et al.~1998),
the primary mechanism that turns quiescent black holes into active galactic
nuclei (AGN) is still debated. 
Galaxy mergers have long been espoused as a possible fueling
mechanism given their effectiveness in dissipating angular momentum and
funneling gas to the center of galaxies (Barnes \& Hernquist 1991; Mihos \& Hernquist 1996).
This can drive both accretion onto the SMBH and growth of the stellar bulge,
which would help explain the tight correlations observed between the two
(e.g., Gebhardt et al.~2000; Ferrarese \& Merritt 2000; Marconi \& Hunt 2003;
H\"{a}ring \& Rix 2004). In fact, recent galaxy merger simulations that 
incorporate a prescription for self-regulated black hole growth have
successfully reproduced many observed properties of AGN and their host galaxies.  
This includes the correlation between SMBH mass and bulge velocity dispersion
(Di Matteo et al.~2007, Robertson et al.~2006), the quasar luminosity
function (Hopkins et al.~2005, 2006a), and the luminosity function of
post-quenched red galaxies (Hopkins et al.~2006b).    
Such simulations have shown that, when coupled with recent AGN feedback
scenarios, major galaxy mergers provide an attractive mechanism to both trigger AGN
activity and help explain the coevolution observed between SMBHs and their
hosts (Hopkins et al.~2008).

Thus far, however, efforts to detect an AGN-merger connection have produced
mixed results.  At low redshifts luminous quasi-stellar objects (QSOs) have
long been tied to ongoing or past merger activity (Stockton 1982; Canalizo \&
Stockton 2001; Bennert et al.~2008).  However Dunlop et al.~(2003) find
that QSOs at $z\sim0.2$ are no more likely to exhibit structural disturbances when compared
to a control sample of similar non-active galaxies.  
At higher redshifts, several studies have used the resolving power of the
\emph{Hubble Space Telescope} (\emph{HST}) to examine the host morphologies of X-ray
selected AGN out to $z\sim1.3$.  
Grogin et al.~(2005) and Pierce et al.~(2007), using data from the
Great Observatories Origins Deep Survey (GOODS, Giavalisco et al.~2004) and
the All-wavelength Extended Groth strip International Survey (AEGIS, Davis et
al.~2007), respectively, find that host galaxies at $z\sim1$ do not show disturbances or
interaction signatures more often than their quiescent counterparts (see also
Sanchez et al.~2004).
More recently, Gabor et al.~(2009) and Cisternas et al.~(2011) examined the
host morphologies of AGN selected in the Cosmic Evolution Survey (COSMOS, Scoville et
al.~2007) and report that the disturbed fraction among active and quiescent
galaxies at $z\sim1$ is not significantly different.  Instead they find that a
majority of AGN at this redshift are hosted by disk galaxies that do not show
strong distortions.  
Schawinski et al.~(2010) recently extended this work to $z\sim2$ by examining
the light profiles of a relatively small number of AGN in a portion of the
GOODS-S field.  They report that a majority of host galaxies at this redshift
have morphologies best fit by low sersic indices indicative of
disk-dominated galaxies and suggest that the bulk of SMBH growth since
$z\sim2$ must be driven by secular processes and not major mergers.  However
this study did not examine the frequency of minor morphological disturbances,
which could be indicative of past interactions, nor did it compare the
morphologies of the AGN hosts against a true mass-matched control sample of
non-active galaxies.   

\begin{figure}[t]
\epsscale{1.1}
\plotone{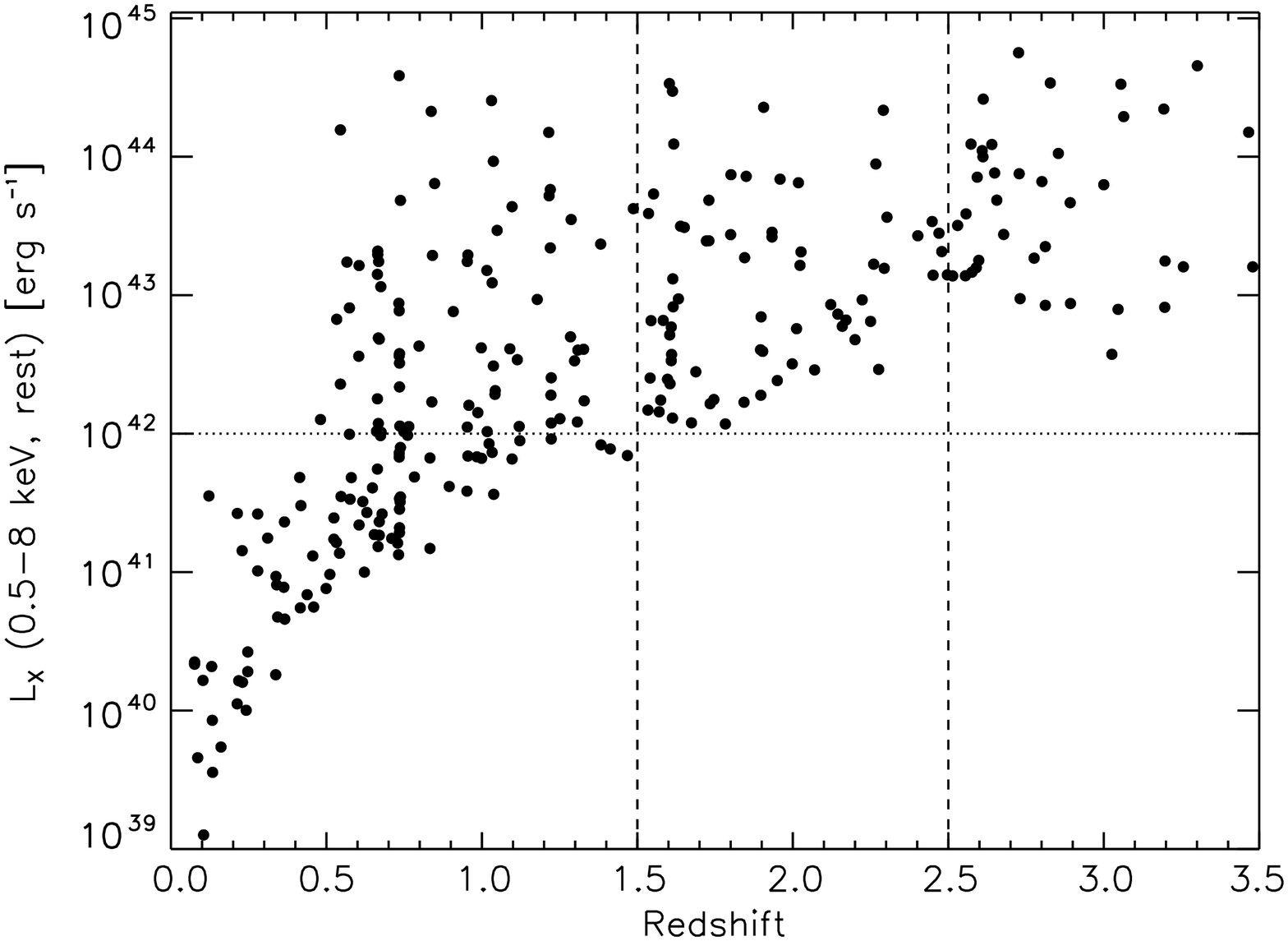}
\caption{\label{fig-z_lum} Redshift versus restframe X-ray luminosity
  (0.5-8 keV) for sources detected in the 4 Msec observations of the CDF-S
  and which fall in the CANDELS and ERS \emph{H}-band
  imaging.  The vertical lines denote the target redshift range of
  $1.5<z<2.5$.  The dotted horizontal line denotes a
  luminosity of $L_{\rm x} = 10^{42}$ erg s$^{-1}$, the maximum X-ray luminosity
  attributable to starburst galaxies. } 
\end{figure}

If there is a redshift at which the primary fueling mechanism of AGN
transitions from secular processes to major mergers, surveys of high
redshift AGN have yet to observe it.   In this study, we extend the search
for an AGN-merger connection for the first time to $z\sim2$, the peak in the
accretion history of the Universe.  
To do this we combine high-resolution near-infrared imaging taken with \emph{HST}/WFC3
as part of the Cosmic Assembly Near-infrared Deep Extragalactic Legacy Survey (CANDELS,
Grogin et al.~2011; Koekemoer et al.~2011) with the 4 Msec \emph{Chandra}
observations of the \emph{Chandra} Deep Field South (CDF-S; Xue et al.~2011),
the deepest X-ray observations obtained to date. 
While \emph{HST}/ACS observations have characterized the restframe ultraviolet
structure of galaxies at $z > 1.5$ (e.g., Jahnke et al.~2004), \emph{HST}/WFC3 observations move beyond the
Balmer break ($\lambda_{\rm rest} \ge 4000$ \AA) and hence probe the light from
stars that dominate a galaxy's mass budget.  This allows us to assess
the rest-frame optical morphologies and true stellar structure of a large
sample of AGN hosts at $z\sim2$ for the first time.
Furthermore, the increased depth of the 4 Msec dataset over previous
\emph{Chandra} observations in the CDF-S provides greater sensitivity to more
obscured (and hence fainter) AGN that may have been missed in previous
studies.

In this study we report on the visual classification of galaxies hosting
X-ray selected AGN in the CDF-S using \emph{HST}/WFC3 imaging in the \emph{H}-band.
We examine whether AGN hosts exhibit an
enhancement of merger or interaction signatures relative to a mass-matched
control sample at the same redshift.  Our analysis is presented in the
following manner: \S2 describes the X-ray and near-infrared data used
for the study, \S3 discusses the AGN sample selection, \S4 details the
morphological classification scheme, and \S5 presents our results.  
Finally, our findings and conclusions are summarized in \S6.  When necessary
the following cosmological parameters are used: $H_{0} = 70 {\rm kms^{-1}
  Mpc^{-1}; \Omega_{tot}, \Omega_{\Lambda}, \Omega_{m} = 1, 0.3, 0.7}$.

\begin{figure}[t]
\epsscale{1.1}
\plotone{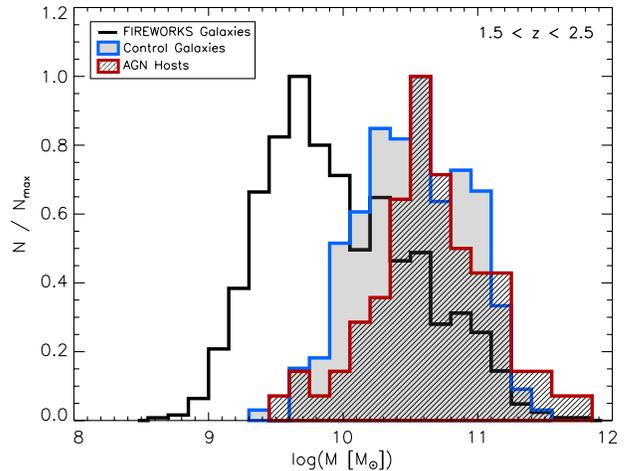}
\caption{\label{fig-masshist} Mass distribution of FIREWORKS galaxies in GOODS-S
  with $K_{s} < 24.3$ AB (\emph{black}) and AGN hosts (\emph{red}) in the redshift range $1.5 < z <
  2.5$.  The distribution of our mass-matched control sample of non-active
  galaxies is shown in blue.}
\end{figure}

\section{Observations and Data Description}

\subsection{Near-Infrared Imaging} 

The near-infrared imaging used for this study consists of WFC3/IR
observations of the GOODS-S field in the F125W ($J$) and F160W ($H$)
bands obtained as part of the Early Release Science program (ERS, Windhorst
et al.~2011) and the ongoing CANDELS Multi-Cycle Treasury Program.  The ERS dataset
consists of 10 WFC3/IR tiles with exposure times of 5000s (each bands)
covering the northern $10^{\prime}\times4^{\prime}$ region of the GOODS-South
field.  The CANDELS dataset consists of 15 WFC3/IR tiles covering the central
$\sim10^{\prime}\times7^{\prime}$ region just south
of and adjacent to the ERS imaging.  At the time of writing, each tile in
the CANDELS dataset has a total exposure time of 3000s in both the $J$ and
$H$ bands.  

The publicly available ERS data and the CANDELS imaging were
reduced as described in Koekemoer et al.~(2011).  Due to the difference
in the exposure times of the observations, they were combined into separate mosaics with matching
pixel scales of 0\farcs{06}/pixel using Multidrizzle (Koekemoer et
al. 2002). From these mosaics we produced an $H$-band selected catalog of
objects in each region using the Source Extractor (SExtractor) software
(Bertin \& Arnouts 1996).  Further details of the CANDELS observations and
data reduction can be found in Grogin et al.~(2011) and Koekemoer et al.~(2011).      

\begin{figure*}[t]
\epsscale{1.1}
\plotone{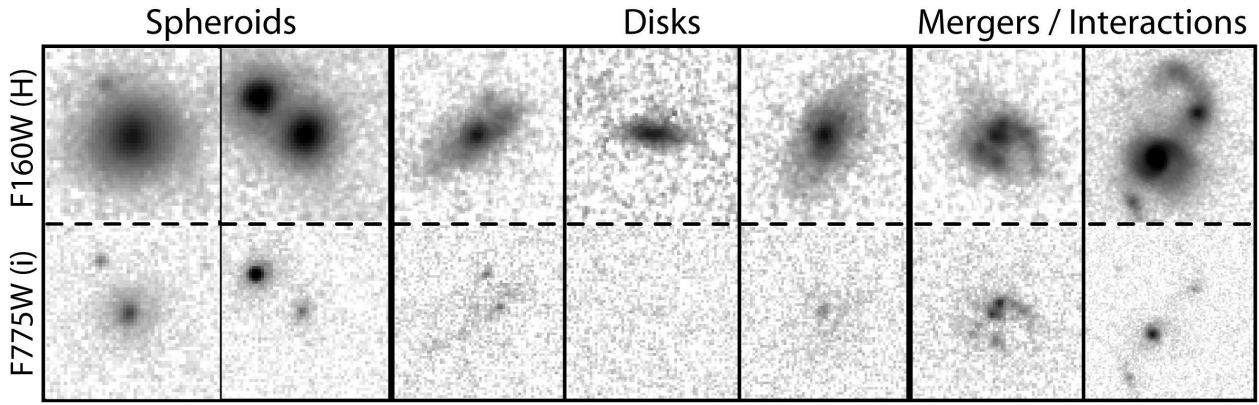}
\caption{\label{fig-montage} Examples of AGN host galaxies that were classified
  as having spheroid and disk morphologies, as well as two galaxies
  experiencing disruptive interactions. Thumbnails on the top row are WFC3/IR images
  taken in the F160W (\emph{H}) band (rest-frame optical), while those on the bottom row are from
  ACS/WFC in the F775W (\emph{i}) band (rest-frame ultraviolet).  These images
demonstrate that accurately classifying the morphology of these galaxies at $z\sim2$ 
requires \emph{H}-band imaging.} 
\end{figure*}

\subsection{X-ray Observations} 

The 4 Msec \emph{Chandra} imaging of the CDF-S was obtained in 54 observations
(obsIDs) over the course of 3 \emph{Chandra} observing cycles in 2000, 2007 and 2010
using the Advanced CCD Imaging Spectrometer imaging array (ACIS-I; Garmire et
al. 2003). The data were reduced using CIAO v4.2 according to the basic
procedure described in Laird et al (2009). Before combining the observations
(and differing from Laird et al), the astrometry of each obsID was registered
to that of the GOODS-MUSYC survey (Gawiser et al 2006) by matching the
positions of bright X-ray sources to H-band selected sources, using the tool
{\tt reproject\_aspect}.  Source detection was carried out according to the method
described in Laird et al (2009). A total of 569 sources were detected to a Poisson
probability limit (i.e.~the wavdetect false-positive threshold) of 4e-6.

\section{Sample Selection}

Our AGN selection is based on X-ray detections in the \emph{Chandra} 4 Msec CDF-S observations.
The power of this dataset to detect AGN at $z\sim2$ stems from its depth
and the fact that at such redshifts hard X-ray emission ($\sim4-5$ keV) is shifted into the
\emph{Chandra} sensitivity window, which peaks at 1.5 keV, potentially allowing us
to detect even heavily obscured SMBHs.  At the target redshift of $z\sim2$, the
luminosity limit of these observations is roughly $L_{\rm x}\sim10^{42}$ erg
s$^{-1}$, the maximum X-ray luminosity thought to be attributable to star
formation processes (Bauer et al.~2002).  This allows us to probe the entire
population of moderate-luminosity AGN at high redshifts for the first
time, while ensuring that the sample is not significantly contaminated by star
forming galaxies without active nuclei. 

To identify $H$-band counterparts to the X-ray sources we employed a maximum likelihood
technique introduced by Sutherland \& Saunders (1992) and described in Brusa
et al.~(2005).  A total of 350/569 X-ray sources in the 4 Msec catalog fall
in the area covered by the ERS and CANDELS imaging.  Of these, 322 were
reliably matched to an $H$-band detected object.  Four spurious
associations are expected among these matches.
Redshifts for the matched hosts were obtained from a variety of publicly
available spectroscopic redshift catalogs (the bulk of these
redshifts come from Silverman et al.~2010) and the
photometric redshifts in the FIREWORKS catalog (Wuyts et al.~2008).  
The latter includes photometric redshift information for 
galaxies with $K_{s} < 24.3$ AB plus detections in at least four additional bands.

Using this combination of redshifts, we find 72 X-ray sources in the target redshift
range of $1.5 < z < 2.5$, of which 22 are new detections not found in the
shallower 2 Msec observations of the CDF-S (Luo et al.~2008).  
The redshift-luminosity distribution of these sources is
shown in Figure 1.  
The sample has a median luminosity of $7.9\times10^{42}$ erg s$^{-1}$ in the
0.5-8 keV band (rest frame).  Five sources have
luminosities in excess of $10^{44}$ erg s$^{-1}$, while the faintest source
has a luminosity of $1.2\times10^{42}$ erg s$^{-1}$. 
Of the 72 \emph{H}-band counterparts, 22
fall within the ERS region, while 50 are located in the CANDELS region.
These 72 galaxies comprise the primary sample of AGN hosts studied in this analysis.

\subsection{Mass-Matched Control Sample}

To properly compare the morphologies of the AGN hosts to similar non-active
galaxies, we constructed a control sample consisting of galaxies with masses
similar to those of the AGN hosts.  Masses for both the active and
non-active populations were obtained from the FIREWORKS catalog, where they
were estimated by modeling each galaxy's observed spectral energy distribution (SED)
with Bruzual \& Charlot (2003) stellar population synthesis models. 
Non-thermal nuclear emission has previously been shown to not significantly
contaminate the rest-frame optical and infrared emission of galaxies hosting
moderate-luminosity AGN such as those in our sample (e.g.~Barger et al.~2005; Bundy et al.~2008).
An examination of the AGN host SEDs confirms this finding for $>90\%$
of the sample.  Therefore we proceed under the assumption that the FIREWORKS-derived
redshifts and masses for the bulk of the AGN sample are not systematically biased
compared to the control sample.  

For each AGN host, we randomly selected three unique, non-active galaxies from the
FIREWORKS catalog whose masses are within a factor of 2 of the AGN host mass
(i.e.,~$M_{\rm AGN}/2 \le M_{\rm gal} \le 2M_{AGN}$).  
Because of the difference in depth between the ERS and CANDELS observations, the
control sample is selected separately for AGN in each region, using only
non-active galaxies in the corresponding region. 
The number of control galaxies was restricted to three per AGN due to the
limited number of massive galaxies available in each field, especially the
smaller ERS region.   

Constructing a mass-matched control sample for this analysis is vital, as
the bulk of the galaxy population at $z\sim2$ is substantially less massive
than the AGN host galaxies.  This can be seen in Figure 2, which shows
the mass distribution for all $K_{s}$-selected FIREWORKS galaxies (with $K_{s} <
24.3$ AB) in the redshift
range $1.5 < z < 2.5$, as well as the masses of the AGN hosts and their
corresponding control galaxies. Without taking mass into consideration, any
control sample selected at $z\sim2$ would be dominated by the low-mass
population, which is predominantly composed of spiral and irregular galaxies,
potentially biasing any morphological comparison.

\section{Visual Classification}

To determine if the AGN host galaxies exhibit merger or interaction
signatures more often than non-active galaxies of similar mass, we have
visually classified the morphologies of both populations.  
Compared to more automated classification techniques, this type of visual
inspection allows us to better pick up low surface brightness features and faint signatures of
past interactions that can be missed using quantitative measures such as
concentration and asymmetry (Kartaltepe et al.~2010).  
Classifiers were asked to determine the predominant
morphological type of each galaxy and the degree to which it was
disturbed.  These inspections were performed blind
(i.e.,~the AGN hosts and control galaxies were mixed) and done primarily in the
$H$-band, although classifiers were also given supplemental $V$, $z$, and
$J$-band images of each galaxy in order to provide additional color information.
The inspectors had access to FITS images in all four bands so as to
manipulate the contrast and stretch of an image if needed. 
Because of the difference in depth of the ERS and CANDELS imaging, the
classifications were carried out separately for the subsamples in each
region.

The classification of each galaxy was split into two categories, with the first
being a general morphological classification and the second a disturbance
classification.  For the former, inspectors were asked to choose among the
following broad morphologies: \emph{Disk, Spheroid, Irregular/Peculiar,
  Point-Like}.  Classifiers were allowed to choose as many as were
applicable to a given galaxy.  For example, a spiral galaxy with a
substantial bulge would be classified as having both a disk and spheroid
component.  

The second category is meant to
gauge the degree to which a galaxy is distorted or disturbed,
presumably as a result of a recent interaction.  The disturbance classes
within this category were designed to pick up not only major mergers, but
also weak interactions that may lead to only minor disturbances or
asymmetries, as well as faint signatures of past merger activity.  Classifiers
were asked to choose one of the following disturbance classes: 

\vspace{0.1in}
\hangindent=0.25in \hangafter=1 $\bullet$ \emph{Merger:} Highly disturbed
with multiple nuclei and/or strong distortions in a single coalescing system. 

\hangindent=0.25in \hangafter=1 $\bullet$ \emph{Interaction:} Two distinct
galaxies showing distortions and interaction features such as tidal arms.

\hangindent=0.25in \hangafter=1 $\bullet$ \emph{Distorted:} Single asymmetric or distorted galaxy with no visible interacting companion. 

\hangindent=0.25in \hangafter=1 $\bullet$ \emph{Double Nuclei:} Multiple nuclei in a single coalesced system.

\hangindent=0.25in \hangafter=1 $\bullet$ \emph{Close Pair:} Near-neighbor
pair in which both are undisturbed. 

\hangindent=0.25in \hangafter=1 $\bullet$ \emph{Undisturbed:} None of the above. \\

While the \emph{Merger} and \emph{Interaction} classes together identify galaxies
in the various stages of a highly disruptive interaction, the
\emph{Distorted} class is meant to flag more subtle signatures of minor
interactions where the companion galaxy may not be visible.  
To improve our statistics, we have grouped several of the interaction classes
into the following categories according to the severity of the observed
disturbance: 

\vspace{0.1in}
\hangindent=0.25in \hangafter=1 $\bullet$ \emph{Disturbed I:} Highly
disturbed systems.  Includes galaxies in the \emph{Merger} and
\emph{Interaction} classes.

\hangindent=0.25in \hangafter=1 $\bullet$ \emph{Disturbed II:} High to
moderately disturbed systems.  Includes all galaxies in the \emph{Disturbed
  I} class, as well as systems with distorted or asymmetric morphologies
(those in the \emph{Distorted} class), and galaxies with \emph{Double Nuclei}. 

\hangindent=0.25in \hangafter=1 $\bullet$ \emph{Companion:} Includes all
galaxies that have a neighbor within 1\farcs{5} (12 kpc projected at
$z\sim2$). This includes all galaxies in the \emph{Close Pair} class.

\vspace{0.1in}

Each galaxy in the sample of 72 AGN hosts and 216 control galaxies was
examined by 12 independent classifiers.  The individual
classifications were combined on a galaxy-by-galaxy basis using the consensus of the group.
For example, a galaxy was assigned a morphological classification only if a
majority (greater than half) of the inspectors agreed that that morphological
component was present.  Multiple morphological classifications per galaxy
were possible.
Since the disturbance classes are mutually exclusive, disturbance
classifications were based on the class most often chosen for a given galaxy.  In
cases where the inspectors were evenly split, the galaxy was assigned the
more disturbed interpretation. For example, a galaxy flagged as an
\emph{Interaction} and as \emph{Disturbed} by an equal number of inspectors
would be assigned the \emph{Interaction} class.  These combined
classifications were then used to calculate the fraction of AGN and control galaxies with a
given morphology or disturbance class.

Further details on the CANDELS visual classification system, including how well the
visual morphologies compare against quantitative morphology measures, can be
found in Kartaltepe et al.~(in preparation).  A particular concern for the AGN host
galaxies is the possibility that nuclear point source emission may mimic a
central bulge component.  We tested for this using parametric Sersic (1968)
fits to the surface brightness profiles of the AGN hosts (van der Wel et
al., in preparation).  The fits were done using the GALFIT package (Peng et
al. 2002) and the GALAPAGOS wrapper.  In general we find broad agreement
between the resulting best-fit Sersic indices, $n$, and the visual morphologies.  
Only a handful of sources show signs of point source contamination, as
evidenced by best-fit Sersic profiles that are steeper than a de Vaucouleurs
profile ($n>4$; de Vaucouleurs 1948).  These sources were predominantly the most
luminous AGN in our sample and were easily identified visually as extended
spheroids, despite the added nuclear emission.  
For these reasons we do not believe that point source contamination has
biased or strongly affecting the visual classification of the bulk of the AGN host sample.

\begin{figure*}[t]
\epsscale{1.0}
\plotone{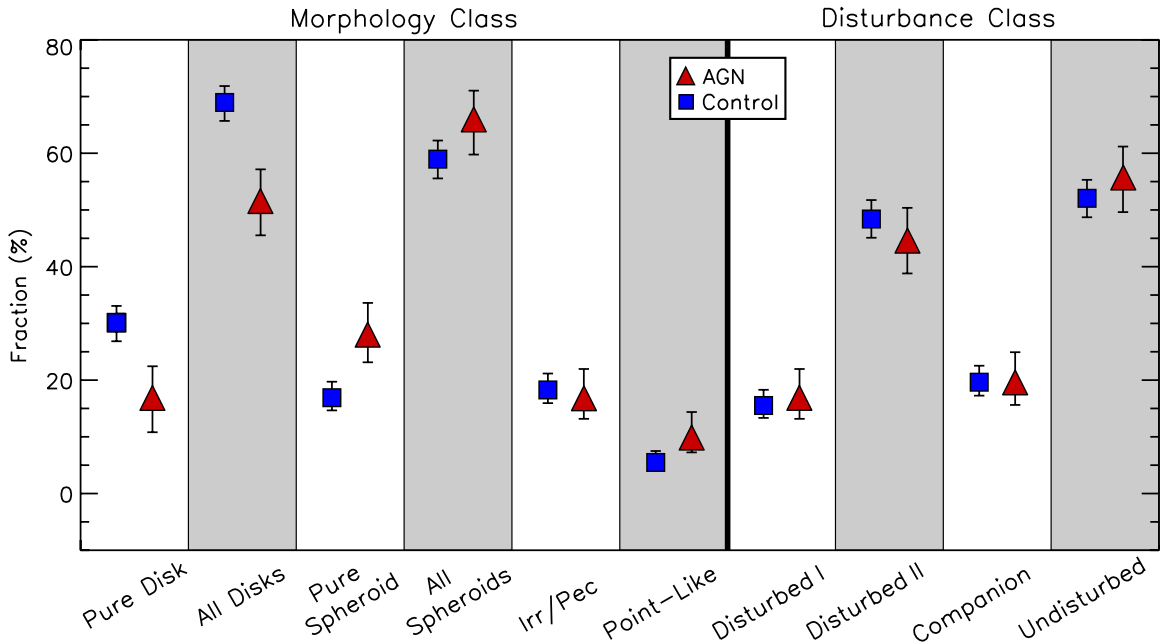}
\caption{Fraction of AGN hosts (\emph{red triangles}) and control galaxies
  (\emph{blue squares}) at $1.5 < z < 2.5$ assigned to
  various morphological and disturbance classes.    The \emph{Pure Disk}
  class includes only disks without a central bulge.  The \emph{Pure Disk} class is a subsample of the
  \emph{All Disks} class, which includes disks with and without a central
  bulge.  Similarly, the \emph{Pure Spheroid} class includes only spheroids
  with no discernible disk component.  The \emph{All Spheroids} class
  includes both \emph{Pure Spheroids} and disk galaxies with a central bulge.
  The \emph{Disturbed I} class is limited to heavily disturbed galaxies in a
  clear merger or interaction.  The \emph{Disturbed II} class includes
  galaxies in the \emph{Disturbed I} class, as well as those showing even
  minor asymmetries in their morphologies\label{fig-results}.  See text for details. }  
\end{figure*}

\section{Results}

Sample WFC3 \emph{H}-band images of AGN host galaxies
that exhibit a range of morphologies and disturbances are shown in Figure
\ref{fig-montage}, while the combined results of the visual analysis of the
AGN hosts and control galaxies are shown in Figure 4 and listed in Table 1.
In the following sections we first present the morphological breakdown of
these galaxies and later discuss the frequency of disturbances observed among them.

A brief note regarding nomenclature: in the following sections the term \emph{disks} (or
\emph{all disks}) refers to all galaxies with a visible disk, including those
with and without a discernible central bulge.  
We will use the term \emph{pure disks} to refer to disk-like galaxies where
no bulge is discernible. On the other hand, \emph{spheroids} (or \emph{pure
  spheroids}) refers to spheroidal galaxies with no discernible disk
component.  At times we will use the term \emph{all spheroids} to refer to both
pure spheroids and the bulge component of galaxies with a visible disk.

\subsection{Host Morphologies}

Shown on the left side of Figure 4 is the fraction of AGN hosts and control
galaxies that were classified as having disk, spheroid, point-like, or
irregular morphologies.  The error bars on each fraction reflect the 68.3\%
binomial confidence limits given the number of sources in each category,
calculated using the method of Cameron et al.~(2010).
For the disk fraction, we show both the fraction of AGN found in pure disks
(i.e.,~those with no discernible bulge) and the fraction hosted by any
disky galaxy (i.e.,~disks with and without a discernible bulge).  

If AGN activity at $z\sim2$ is triggered predominantly by major
mergers, we might expect an increased incidence of irregular morphologies
among the AGN hosts.  Instead, we find the irregular fraction to be relatively
low ($16.7^{+5.3}_{-3.5}\%$) and consistent with the fraction observed among the control
sample ($18.2^{+2.9}_{-2.3}\%$).  This is the first indication that AGN are not found in
substantially disturbed galaxies more often than their non-active
counterparts at this redshift.  In fact, a high fraction of the AGN are found in galaxies
with a visible disk, a component which is unlikely to have survived a major merger
in the recent past.  We find disks to be the most common single morphology assigned
to the AGN hosts, making up $51.4^{+5.8}_{-5.9}\%$ of the entire sample.  
Two-thirds of these galaxies (67.6\%) also exhibit a prominent bulge
component, while 32.4\% of the disk galaxies show no discernible central
bulge. Of the remaining hosts, pure spheroids comprise $27.8^{+5.8}_{-4.6}\%$ of the entire sample,
while point-like sources constitute $9.7^{+4.7}_{-2.5}\%$.

Despite the prevalence of disks among the AGN hosts, we find that the active
galaxies are more often associated with spheroid morphologies than their
non-active counterparts. 
Pure spheroids make up $27.8^{+5.8}_{-4.6}\%$ of the AGN hosts versus only
$16.9^{+2.8}_{-2.2}\%$ of the massive control galaxies.  Disks, on the other hand,
are more common among the control sample, comprising $69.0^{+2.9}_{-3.3}\%$
of the non-active galaxies, but only $51.4^{+5.8}_{-5.9}\%$ of the AGN hosts.
The smaller disk fraction among the active galaxies compared to the control
galaxies is significant at the 99.6\% level, assuming a binomial error
distribution.   

The fact that the AGN tend to favor more spheroid-dominated hosts is further
illustrated in Figure 5, where we show the fraction of active and control
galaxies classified as pure disks, disks with central bulges and pure
spheroids.  The AGN host morphologies are skewed toward more
spheroid-dominated systems as they show an excess of pure spheroid
morphologies relative to the control sample and a deficit of pure disk
morphologies.  We find that bulgeless, pure disks
constitute $30.1^{+3.3}_{-2.9}\%$  of the control population while making up
only $16.7^{+5.3}_{-3.5}\%$ of the AGN host galaxies.  These findings
suggests that the trend observed at lower redshifts, that AGN hosts are more
spheroid-dominated relative to similarly massive non-active galaxies
(e.g.,~Grogin et al.~2005; Pierce et al.~2007), continues to some extent out
to $z\sim2$. 

Lastly, we have considered the possibility that host morphology, and hence
triggering mechanisms, vary systematically with X-ray luminosity.
To investigate this, we have examined the morphologies of active galaxies
with X-ray luminosities above and below $L_{\rm X} = 10^{43}$ erg s$^{-1}$.
These subsamples includes 32 and 40 AGN, respectively, out of the full sample of 72 at $z\sim2$.
The morphological breakdown of these subsamples is listed in Table 1.  
We find no increase in the irregular fraction among the more luminous AGN, but we
do observe a dramatic reversal in the spheroid and disk fractions: spheroids
constitute $40.6^{+9.0}_{-7.9}\%$ of the galaxies hosting the more X-ray
luminous AGN, while the disk fraction drops to $34.4^{+9.1}_{-7.3}\%$.  This
is compared to a spheroid and disk fraction of $18.4^{+7.9}_{-4.7}\%$ and
$68.4^{+6.5}_{-8.3}\%$, respectively, for the lower luminosity sample.  This
finding agrees with local host properties, where luminous AGN and QSOs tend
to be more often associated with early-type hosts (Kauffmann et al.~2003).

\subsection{Interaction Signatures}

The fraction of active and non-active galaxies that exhibit various levels of
disturbance in their morphologies is shown on the right side of Figure 4 and
the bottom part of Table 1.
We find that $16.7^{+5.3}_{-3.5}\%$ of the AGN hosts are involved in highly disruptive
mergers or interactions and fall in the \emph{Disturbed I} category.
This percentage is statistically no different than the fraction of similarly
disturbed non-active galaxies ($15.5^{+2.8}_{-2.2}\%$).
If we include galaxies showing minor asymmetries
in their morphologies and those with double nuclei (i.e.,~\emph{Disturbed II}
systems), the fraction of disturbed active galaxies increases to $44.4^{+5.9}_{-5.6}\%$.  This
is below the percentage of control galaxies that fall in the same category
($48.4^{+3.4}_{-3.4}\%$), but the fractions are again statistically equivalent.   
In fact, for all of the distortion classes we considered, the properties of
the AGN hosts are not significantly different than the non-active sample.
This includes the fraction showing clear merger and interaction signatures
(see Table 1), those with only a minor asymmetry in their morphology
($30.6^{+4.9}_{-5.9}\%$ for AGNs versus $32.9^{+3.0}_{-3.3}\%$
for the control sample, respectively), and the frequency of companions
within 1\farcs{5} (12 kpc projected; $19.4^{+5.5}_{-3.8}\%$ versus $19.6^{+3.0}_{-2.4}\%$). 

The most common disturbance class assigned to both the AGN hosts and control
galaxies is \emph{Undisturbed}, making up $55.6^{+5.6}_{-5.9}\%$ and $52.1^{+3.3}_{-3.4}\%$ of each population,
respectively.  This suggests that a majority of the AGN at $z\sim2$ reside in
relatively relaxed galaxies that do not show even minor disturbances in our
$H$-band imaging.  Significantly, these results do not change when we limit our analysis to
the more luminous AGN in the sample: the \emph{Undisturbed} fraction
is still $46.9^{+8.7}_{-8.4}\%$ even when only AGN with  $L_{\rm X} >10^{43}$
erg s$^{-1}$ are considered.  Furthermore, the fraction of hosts in the
\emph{Disturbed I} and \emph{Disturbed II} categories are $18.8^{+8.7}_{-5.0}\%$ and $53.1^{+8.4}_{-8.8}\%$,
respectively, and in rough agreement with the percentages found for the full
sample.

\begin{center}
\tabletypesize{\scriptsize}
\begin{deluxetable*}{ccccc}
\tablewidth{0pt}
\tablecaption{Visual Classification Results}
\tablecolumns{5}
\tablehead{\colhead{} & \colhead{AGN} & \colhead{Control} & \colhead{AGN} & \colhead{AGN} \\  
           \colhead{Classification} & \colhead{Hosts} & \colhead{Galaxies} &
           \colhead{$L_{\rm X} < 10^{43}$ erg s$^{-1}$} & \colhead{$L_{\rm X} > 10^{43}$ erg s$^{-1}$} }
\startdata

Pure Disk        & $16.7^{+5.3}_{-3.5}\%$  &  $30.1^{+3.3}_{-2.9}\%$   &  $21.0^{+8.0}_{-5.1}\%$ &  $12.5^{+8.2}_{-3.7}\%$     \\ 
All Disks        & $51.4^{+5.8}_{-5.9}\%$  &  $69.0^{+2.9}_{-3.3}\%$   &  $68.4^{+6.5}_{-8.3}\%$ &  $34.4^{+9.1}_{-7.3}\%$     \\ 
Pure Spheroid    & $27.8^{+5.8}_{-4.6}\%$  &  $16.9^{+2.8}_{-2.2}\%$   &  $18.4^{+7.9}_{-4.7}\%$ &  $40.6^{+9.0}_{-7.9}\%$     \\ 
All Spheroids    & $62.5^{+5.3}_{-6.0}\%$  &  $55.7^{+3.3}_{-3.4}\%$   &  $65.8^{+6.7}_{-8.3}\%$ &  $62.5^{+7.5}_{-9.1}\%$     \\ 
Irregular        & $16.7^{+5.3}_{-3.5}\%$  &  $18.2^{+2.9}_{-2.3}\%$   &  $21.0^{+8.0}_{-5.1}\%$ &  $06.3^{+7.3}_{-2.1}\%$     \\ 
Point-like       & $09.7^{+4.7}_{-2.5}\%$  &  $05.5^{+2.0}_{-1.2}\%$   &  $02.6^{+5.6}_{-0.8}\%$ &  $18.8^{+8.7}_{-5.0}\%$     \\ 
\hline                                                                                           
Disturbed I       & $16.7^{+5.3}_{-3.5}\%$  &  $15.5^{+2.8}_{-2.2}\%$  &  $15.8^{+7.7}_{-4.2}\%$ &  $18.8^{+8.7}_{-5.0}\%$     \\  
Disturbed II      & $44.4^{+5.9}_{-5.6}\%$  &  $48.4^{+3.4}_{-3.4}\%$  &  $36.8^{+8.3}_{-7.0}\%$ &  $53.1^{+8.4}_{-8.8}\%$     \\  
Companion         & $19.4^{+5.5}_{-3.8}\%$  &  $19.6^{+3.0}_{-2.4}\%$  &  $18.4^{+7.9}_{-4.7}\%$ &  $21.9^{+8.9}_{-5.6}\%$    \\
Undisturbed       & $55.6^{+5.6}_{-5.9}\%$  &  $52.1^{+3.3}_{-3.4}\%$  &  $63.2^{+7.0}_{-8.3}\%$ &  $46.9^{+8.7}_{-8.4}\%$     \\    

\vspace*{-0.075in}
\enddata
\tablecomments{The \emph{Pure Disk} and \emph{Pure Spheroid} classes are
  included in the \emph{All Disks} and \emph{All Spheroids} classes, respectively. 
  Likewise, the \emph{Disturbed I} class is a subset of the \emph{Disturbed II} class.}
\end{deluxetable*}
\end{center}

\vspace{-0.3in}
\section{Discussion}

In summary, our primary findings are: i)
moderate luminosity, X-ray selected AGN at $z\sim2$ do not exhibit a
significant excess of distorted morphologies relative to a mass-matched
control sample at the same redshift; ii) both samples are dominated by
systems that appear relatively relaxed and undisturbed, to the depth of our
imaging; and iii) a large fraction ($51.4^{+5.8}_{-5.9}\%$) of the AGN reside
in galaxies with discernible disks.  
Therefore, based on our visual classifications, we do not find a strong
connection between highly disruptive major mergers and moderate luminosity
AGN activity at $z\sim2$.  

If mergers play an important role in triggering AGN activity, there are two
possible effects that have been discussed in the literature which could help
explain the lack of disturbed morphologies among the AGN hosts.  The first is
obscuration; if obscured AGN are preferentially associated with mergers, they
may be systematically missed by X-ray surveys.  It is well known
that, in the local universe, gas rich mergers have extremely high dust column
densities, which may be sufficient to hide even hard X-ray sources deep in
the nuclei (Hopkins et al. 2007).  The second is a time delay between the
onset of AGN activity and the actual merger.  If this delay is of the same
order as the relaxation time of the galaxy (typically $\sim$few 100 Myr), the most
obvious signatures of morphological disturbance will have faded by the time
it is identified as an X-ray bright AGN (Lotz et al.~2010).  Hydrodynamic
simulations of SMBH growth in galaxy mergers do predict such a delay (Hopkins
et al.~2006), and there is observational evidence for a delay of order
$\sim$250 Myr between starburst and AGN activity (Wild, Heckman, \& Charlot 2010). 

While these two effects can help explain the lack of obvious merger
signatures among the AGN hosts, the high disk fraction we observe is harder to
reconcile with the merger picture of AGN fueling. This finding that disk-like
morphologies are prevalent among active galaxies at this redshift agrees with the
recent findings of Schawinski et al.~(2011), 
who examined a smaller sample of AGN at $z\sim2$ in the ERS region of
GOODS-South.   This characteristic does not appear to be limited to the AGN
hosts, though, as we find disks are common among all massive galaxies at this
redshift, regardless of whether they host an active nucleus.  
Nevertheless, the morphology of these galaxies provides an important clue to
the mechanism that triggered their current AGN activity.
It is doubtful that the disk-like structure of these galaxies could have survived the
large-scale and violent torquing of gas that occurs during a \emph{major} galaxy-galaxy
interaction (e.g.~Bournaud et al.~2011), making it highly unlikely that their nuclear activity is being
fueled by a major merger-driven process.
It is more likely that the nuclear activity in these disk galaxies is being
fueled by the stochastic accretion of cold gas, possibly triggered by a
disk instability or minor interaction that did not substantially perturb the
large-scale structure of the galaxy.

Despite the prevalence of disk-like morphologies, we also find the AGN are more often
associated with spheroids than their non-active counterparts.  While the connection between
spheroid-dominated galaxies and AGN has been well established at lower
redshifts, this is the first such finding at $z\sim2$, where there is evidence that
the canonical mass-morphology relationship appears to break down (McGrath et
al.~2008; van der Wel et al.~2011).  Even in an era where massive galaxies
predominately have disk-like morphologies, our observations suggest SMBH
continue to be preferentially embedded in spheroidal systems.  

\begin{figure}[t]
\epsscale{1.1}
\plotone{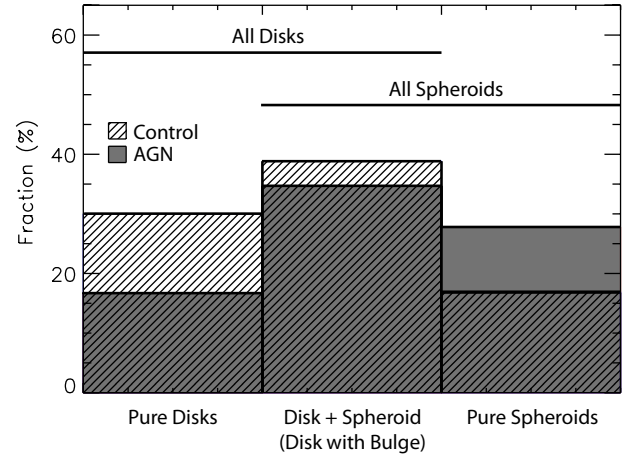}
\caption{Fraction of AGN hosts and control galaxies at $1.5 < z < 2.5$ classified as
bulgeless pure disks, disks with central bulges and pure spheroids.  The AGN
tend to favor more spheroid-dominated hosts as they show an excess of pure spheroid
morphologies relative to the control sample and a deficit of pure disk
morphologies. \label{fig-agnsph} }  
\end{figure}

At higher X-ray luminosities ($L_{\rm X}>10^{43}$ erg s$^{-1}$) we also
observe a shift in the overall AGN host population toward spheroids.
Although many of these galaxies appear undisturbed, they may have been
triggered by a merger event in the recent past and have since sufficiently
relaxed.  In the evolutionary sequence of a merger-triggered QSO presented in
Hopkins et al.~(2007), these systems would be in the post-blowout phase, when
nuclear activity and X-ray luminosity are in gradual decline.  

Our finding that more luminous AGN are more often associated with spheroids
generally agrees with the luminosity and redshift 
dependent AGN fueling model presented by Hopkins \& Hernquist (2006; hereafter HH06), which
proposes that luminous AGN and QSOs are largely triggered by major mergers, 
while lower luminosity AGN are fueled by the random accretion of
gas via internal, secular processes. 
However this model also predicts that by
$z\sim2$ the active galaxy population should be dominated by merger-triggered
AGN, even at moderate luminosities ($L_{\rm x} \sim 10^{43}$ erg s$^{-1}$).
This is because merger-driven accretion in this model is tied to the
cosmological galaxy merger rate, which increases rapidly with redshift
(Conselice et al.~2003; Kartaltepe et al.~2007) whereas quiescent accretion
is related to the mass function and gas fraction of late-type galaxies, which
evolve more slowly.    

The HH06 model predicts that at $z=2$ the number density of quiescently
accreting AGN will not equal that of merger-fueled AGN until roughly \emph{two orders of
magnitude} below the knee in the AGN luminosity function.  In the hard
X-ray band this knee occurs at a luminosity of roughly $L_{\rm X} \sim
10^{44}$ erg s$^{-1}$ (Aird et al.~2010).    
This means the predicted X-ray luminosity at which an equal fraction of AGN are
fueled by quiescent and merger-triggered accretion at $z=2$ is roughly $L_{\rm
  X} \sim 10^{42}$ erg s$^{-1}$.  
If we assume that disk-like hosts are fueling their AGN via internal processes
and have not experienced a major merger in the recent past, then this
prediction is at odds with the high disk fraction we observe at  $L_{\rm X} \sim
10^{43}$ erg s$^{-1}$, an order of
magnitude above this luminosity.  
We find that the luminosity at which an equal fraction of AGN
are hosted by disk and spheroid galaxies is roughly $L_{\rm X} \sim 10^{43}$ erg s$^{-1}$.
This finding suggests that the stochastic fueling of SMBHs is far more
prevalent at moderate luminosities than predicted by the HH06 model. 

This apparent disagreement with the HH06 fueling model
was previously reported at lower redshifts by Georgakakis et al.~(2009), who found
that the contribution to the X-ray luminosity function at $z\sim1$ from AGN in late-type hosts exceeded the
predicted luminosity function for stochastically fueled AGN.  
It was also noted by Cisternas et al.~(2011), who found a large fraction
(55.6\%) of luminous AGN ($L_{\rm X} > 10^{44}$ erg s$^{-1}$) at $z\sim1$
hosted by disk-dominated galaxies. 
While the high disk fraction we observe is similar to what has been
previously reported in these studies, the disagreement between our findings and the predictions
of the HH06 model is more acute given the
higher redshift of our sample and the strong redshift evolution predicted for
merger-driven accretion.  

Overall our findings generally agree with an emerging consensus that major
galaxy mergers likely play a subdominant role in triggering moderate-luminosity AGN.
This has been asserted from a morphological standpoint 
by Cisternas et
al.~(2011) and Georgakakis et al.~(2009) at $z\sim1$, and by Schawinski et
al.~(2011) at $z\sim2$, based on the large disk fraction found among AGN hosts.
It has also been proposed by Mullaney et al.~(2011) based on the average
specific star formation rates (SSFR) of AGN hosts out to $z\sim3$.  They find that a
vast majority of hosts have SSFRs consistent with the star forming main
sequence (Noeske et al.~2007) and that less than 10\% appear to be undergoing a starburst
phase.  From this they conclude that the nuclear activity in these galaxies
is being fueled by internal mechanisms rather than violent mergers.  
A similar conclusion was also reached by Allevato et al.~(2011) based on the
projected clustering of AGN in the COSMOS field out to $z\sim2.2$.

There are several reasons why non-merger related accretion may contribute
more to the onset of AGN activity at this redshift than previously expected.  
This includes such things as a shorter post-blowout quasar lifetime, which
would reduce the contribution from merger-triggered AGN to the X-ray
luminosity function, or a faster evolving gas fraction than that assumed by
HH06. It may also be due to the rise of violent
gravitational instabilities in disk galaxies due to the effects of rapid cold
flow accretion (Dekel, Sari, \& Ceverino 2009).  Such instabilities become
increasingly common at $z>1$ (Elmegreen et al.~05, Genzel et al.~06) and are
not accounted for in the HH06 model.  Unlike the
weaker disk instabilities that are associated with secular evolution at low
redshift (e.g.,~bar instabilities), these high redshift instabilities are
highly efficient at continuously funneling gas and stars to the centers of
galaxies on short timescales (a single disk rotation) and at high inflow
rates ($\sim10 M_{\odot}$ yr$^{-1}$; Cacciato \& Dekel 2011), potentially
fueling increased AGN activity in disk galaxies without the need for
galaxy-galaxy mergers (Bournaurd et al.~2011, in prep).  

Of course, the disagreement between the high disk fraction we
observe and the merger-dominated fueling model is predicated on the
assumption that disk-like hosts have not experienced a merger in the recent past. 
The two can be reconciled if these disks have instead survived or reformed
following a merger event.  Numerical simulations have shown that disks can
reform after a merger if the interacting systems are gas rich (Robertson et
al.~2006; Bundy et al.~2010), although it has been argued that such interactions are not
conducive to the fueling of SMBHs (Hopkins \& Hernquist 2009).  
Alternatively, minor mergers provide a means to trigger AGN-activity within
galaxies without entirely destroying their pre-existing morphology.  
Semi-analytic cosmological galaxy formation models in which all AGN activity
is assumed to be triggered by mergers (Somerville et al. 2008) do predict that
the average merger event that triggers an AGN with $L_{\rm X} > 10^{42}$ erg s$^{-1}$ has a mass
ratio of 1:8 as opposed to the more disruptive 1:1 or 1:2 mergers (Somerville et al.~2011). 
Coupled with a time delay between the merger and the visibility of the AGN,
the signatures of these mergers could prove difficult to detect.
Therefore, since we cannot rule out such interactions, minor mergers would
seem to be one of the remaining ways to reconcile the merger-dominated
fueling model with the high disk fraction and lack of disturbed morphologies
that we observe.

\section{Conclusions}

To explore if major galaxy mergers are the primary mechanism fueling AGN
activity at $z\sim2$, we have used \emph{HST}/WFC3 imaging to examine the rest-frame optical morphologies
of galaxies hosting moderate-luminosity, X-ray selected AGN at $z=1.5-2.5$.
Employing visual classifications, we have determined both the predominant
morphological type of these galaxies and the frequency at which they exhibit
morphological disturbances indicative of recent interactions.
To determine if the AGN hosts show merger or interaction
signatures more often than similar non-active galaxies, we have also classified a
sample of mass-matched control galaxies at the same redshift.

First, we find that just over half of the AGN reside in disk galaxies
($51.4^{+5.8}_{-5.9}\%$), while a smaller percentage are found in spheroids ($27.8^{+5.8}_{-4.6}\%$)
and systems with irregular morphologies ($16.7^{+5.3}_{-3.5}\%$).  
This high disk fraction is also observed among the control galaxies
($69.0^{+2.9}_{-3.3}\%$ of which are disks), which indicates disk-like
morphologies are prevalent among all massive galaxies at this redshift,
regardless of whether they host an active nucleus.
In fact, we find the AGN to be more often associated with spheroidal galaxies
compared to their non-active counterparts.  Pure spheroids account for
$27.8^{+5.8}_{-4.6}\%$ of the active galaxies, while comprising only
$16.9^{+2.8}_{-2.2}\%$ of the control galaxies. At X-ray
luminosities above 10$^{43}$ erg s$^{-1}$ we observe a reversal in the
morphological make-up of the AGN hosts, with spheroids and disks
comprising $40.6^{+9.0}_{-7.9}\%$ and $34.4^{+9.1}_{-7.3}\%$  of the sample, respectively.

Second, we find that $16.7^{+5.3}_{-3.5}\%$ of the AGN hosts have highly disturbed
morphologies and appear to be involved in a major merger or interaction,
while $44.4^{+5.9}_{-5.6}\%$  show at least some disturbance, including minor asymmetries in
their morphologies. 
In both cases, these fractions are statistically consistent with the fraction of control galaxies
that show similar morphological disturbances.   Most galaxies in both the
control and AGN samples appear relatively relaxed and undisturbed, to the
depth of our imaging ($55.6^{+5.6}_{-5.9}\%$ and $52.1^{+3.3}_{-3.4}\%$,
respectively).  These results suggest
the AGN hosts are no more likely to be involved in an ongoing major merger or
interaction than non-active galaxies of similar mass. 

Finally, the high disk fraction observed among the AGN hosts appears to be at
odds with predictions that major merger-driven accretion should be the dominant
AGN fueling mode at $z\sim2$.  
The presence of a large population of relatively undisturbed disk-like hosts
suggests that either secular evolution, disk instability-driven accretion,
minor mergers, or a combination of the three, play a greater role in
triggering AGN activity at these redshifts than previously thought.  
In a forthcoming paper
we plan to calculate the fraction of the AGN X-ray luminosity function
attributable to these disk-hosted AGN in order to quantify any discrepancy
between our observations and merger-triggered AGN fueling models.

\vspace{0.5in}
Support for Program number HST-GO-12060  was provided by NASA through a grant from the Space Telescope Science Institute, which is operated by the Association of Universities for Research in Astronomy, Incorporated, under NASA contract NAS5-26555.
Furthermore, D.K. is funded in part by the NSF under Grant No. AST-0808133.

\bibliography{ms.revised_eapj.bbl}

\end{document}